\documentclass{revtex4-1}
\usepackage{graphicx}
\usepackage{color}
\usepackage{subfigure}
\begin{document}
%\linenumbers

%\def\rmR{\mbox{\scriptsize{R}}}
%\def\rmrel{\mbox{\scriptsize{rel}}}
%\def\{\tt WCC}{$W_{\mbox{\scriptsize{cc}}}$}
%\def\{\tt WCR}{$W_{\mbox{\scriptsize{cr}}}$}
%\def\{\tt EABS}e{$E_{\mbox{\scriptsize{abs}}}^{e^{\pm}}$}
%\def\{\tt EABS}g{$E_{\mbox{\scriptsize{abs}}}^{\gamma}$}
%\def\Ncompton{$N_{\mbox{\scriptsize{Co}}}$}
%\def\EPSF{$\left< E \right>_{\mbox{\scriptsize{PSF}}}$}

%\title{A geometry for the Monte Carlo simulation of TrueBeam linacs}
\title{A geometrical model for the Monte Carlo simulation of the TrueBeam linac}

\author{Miguel Rodriguez}
\affiliation{Institut de T\`{e}cniques Energ\`{e}tiques, Universitat Polit\`{e}cnica de Catalunya \\ Diagonal 647, E-08028 Barcelona, Spain.}

\author{Josep Sempau}
\affiliation{Institut de T\`{e}cniques Energ\`{e}tiques, Universitat Polit\`{e}cnica de Catalunya \\ Diagonal 647, E-08028 Barcelona, Spain.}

\author{Antonella Fogliata}
\affiliation{Radiotherapy and Radiosurgery Department, Humanitas Clinical and Research Center\\ Milan-Rozzano, Italy.}

\author{Luca Cozzi}
\affiliation{Radiotherapy and Radiosurgery Department, Humanitas Clinical and Research Center\\ Milan-Rozzano, Italy.}

\author{Wolfgang Sauerwein}
\affiliation{NCTeam, Strahlenklinik, Universit\"atsklinikum Essen,\\ Hufelandstra\ss e 55 D-45122 Essen, Germany.}

\author{Lorenzo Brualla}
\email{lorenzo.brualla@uni-duisburg-essen.de}
\thanks{Corresponding author}
\homepage[This is an author-created, un-copyedited version of an article accepted for publication in Physics in Medicine and Biology. IOP Publishing Ltd is not responsible for any errors or omissions in this version of the manuscript or any version derived from it. The definitive publisher authenticated version (Phys. Med. Biol. 60 (2015) N219--N229) is available online at: ]{http://dx.doi.org/10.1118/1.4916686}
\affiliation{NCTeam, Strahlenklinik, Universit\"atsklinikum Essen,\\ Hufelandstra\ss e 55 D-45122 Essen, Germany.}

\begin{abstract}
Monte Carlo simulation of linear accelerators (linacs) depends on the accurate geometrical description of the linac head. The geometry of the Varian TrueBeam linac is not available to researchers. Instead, the company distributes phase-space files of the flattening-filter-free (FFF) beams tallied at a plane located just upstream the jaws. Yet, Monte Carlo simulations based on third party tallied phase spaces are subject to limitations. In this work, an experimentally-based geometry developed for the simulation of the FFF beams of the Varian TrueBeam linac is presented. The Monte Carlo geometrical model of the TrueBeam linac uses information provided by Varian which reveals large similarities between the TrueBeam machine and the Clinac 2100 downstream of the jaws. Thus, the upper part of the TrueBeam linac was modeled by introducing modifications to the Varian Clinac 2100 linac geometry. The most important of these modifications is the replacement of the standard flattening filters by {\it ad hoc} thin filters. These filters were modeled by comparing dose measurements and simulations. The experimental dose profiles for the 6~MV and 10~MV FFF beams were obtained from the Varian Golden Data Set and from in-house measurements performed with a diode detector for radiation fields ranging from $3\times3$ to $40\times40$ cm$^2$ at depths of maximum dose, 5 and 10~cm. Indicators of agreement between the experimental data and the simulation results obtained with the proposed geometrical model were the dose differences, the root-mean-square error and the gamma index. The same comparisons were done for dose profiles obtained from Monte Carlo simulations using the phase-space files distributed by Varian for the TrueBeam linac as the sources of particles. Results of comparisons show a good agreement of the dose for the ansatz geometry similar to that obtained for the simulations with the TrueBeam phase-space files for all fields and depths considered, except for the $40\times40$ cm$^2$ field where the ansatz geometry was able to reproduce the measured dose more accurately. Our approach overcomes some of the limitations of using the Varian phase-space files. It makes possible to: (i) adapt the initial beam parameters to match measured dose profiles; (ii) reduce the statistical uncertainty to arbitrarily low values; and (iii) assess systematic uncertainties (type B) by employing different Monte Carlo codes. One limitation of using phase-space files that is retained in our model is the impossibility of performing accurate absolute dosimetry simulations since the geometrical description of the TrueBeam ionization chamber remains unknown.

\end{abstract}

%\pacs{00.00, 00.00, 00.00}

%\noindent{\it Keywords}: Monte Carlo, {\sc penelope}, linear accelerator, TrueBeam

%\vspace{5pc}\noindent

%\submitto{\PMB}

\maketitle

%==================== Introduction ====================%

\section{Introduction}\label{sec:Intro}

The TrueBeam linac is the most recently marketed linear accelerator by Varian Medical Systems (Palo Alto, California, USA). The TrueBeam linac can produce flattening-filter-free (FFF) photon beams of 6 and 10~MV. Removing the flattening filter is an effective way to boost the dose rate in high-dose treatments with multiple narrow fields such as those employed in stereotactic ablative radiotherapy.

Since the seminal works of Petti and co-workers~\cite{Petti1983} and of Mary Udale~\cite{Udale1988} on Monte Carlo simulation of a linac until more recent times (see~\cite{Verhaegen2003,Chetty2006} and references therein) researchers have tried to improve the accuracy with which Monte Carlo models reproduce experimental dose distributions in water and heterogeneous phantoms. It has become clear that this goal requires precise knowledge of the dimensions and materials of the linac head assembly components that influence the beam. Usually, these data are provided by the manufacturers to Monte Carlo users who agree to keep this information undisclosed. However, in the case of the TrueBeam linac, Varian has not made available the characteristics of the head components situated above the collimation jaws. Instead, the company distributes, through its website (http://www.myvarian.com/montecarlo), phase-space files of the FFF beams tallied at a plane located just upstream the jaws which can be downloaded by TrueBeam users. These phase-space files can be used as sources to transport particles through the geometry of the jaws, and other beam modifiers, for subsequently estimating the absorbed dose in a patient computerized tomography or phantom. Those phase-space files were produced by Varian \cite{Constantin2011} in simulations employing the Monte Carlo code Geant4 \cite{Geant4} and are coded in the International Atomic Energy Agency (IAEA) phase-space format \cite{Capote2006}.

Some authors have validated the Varian distributed phase-space files and they have generally found a reasonable agreement of the estimated dose with measurements \cite{Constantin2011,Gete2013,Belosi2014}. However, Monte Carlo simulations based on third party tallied phase spaces are subject to several limitations: (i) the initial beam parameters cannot be adapted to match dose profiles measured in the user's linac; (ii) the latent variance \cite{Sempau2001} of the phase-space file imposes a limit in the statistical uncertainty of the estimated observables; (iii) simulations are subject to type B uncertainties associated to limitations of the Monte Carlo code used for generating the phase-space files \cite{Faddegon2009}; and (iv) the energy deposited in the ionization chamber cannot be determined since this component is located in the upper part of the linac, thus, absolute dosimetry in units of Gy/MU cannot be accurately computed by the method of Popescu {\it et. al} \cite{Popescu2005}.

In this work we present a Monte Carlo model of the TrueBeam linac, inferred by comparing dose measurements and simulations, which is suitable for the simulation of the 6 and 10 MV FFF beams. The geometry of the linac in this model includes thin filters for the 6 and 10~MV beams that are situated approximately in the position occupied by the standard flattening filters of a Varian Clinac 2100 linac. We have called this geometry FakeBeam to emphasize that it is not the actual geometry of the TrueBeam machine, but an {\it ad hoc} one created with the purpose of providing Monte Carlo users a way to circumvent some of the aforementioned limitations of using the Varian distributed phase-space files.

\section{Materials and methods}\label{sec:M&M}
\subsection{FakeBeam geometry}\label{sec:FB}
Simply removing the flattening filters of a Clinac~2100 linac is not enough to reproduce the dose distributions produced by the TrueBeam FFF beams in a water phantom. The sole elimination of the flattening filter would allow high-energy electrons escaping from the target to produce a considerably higher than expected dose at shallow depths. The insertion of a thin sheet of a material such as bronze or copper at some place in the beam path can absorb those contaminant electrons. However, this simple configuration still fails to reproduce the off-axis absorbed dose.

\begin{figure*}
\begin{center}
\includegraphics[scale=0.75]{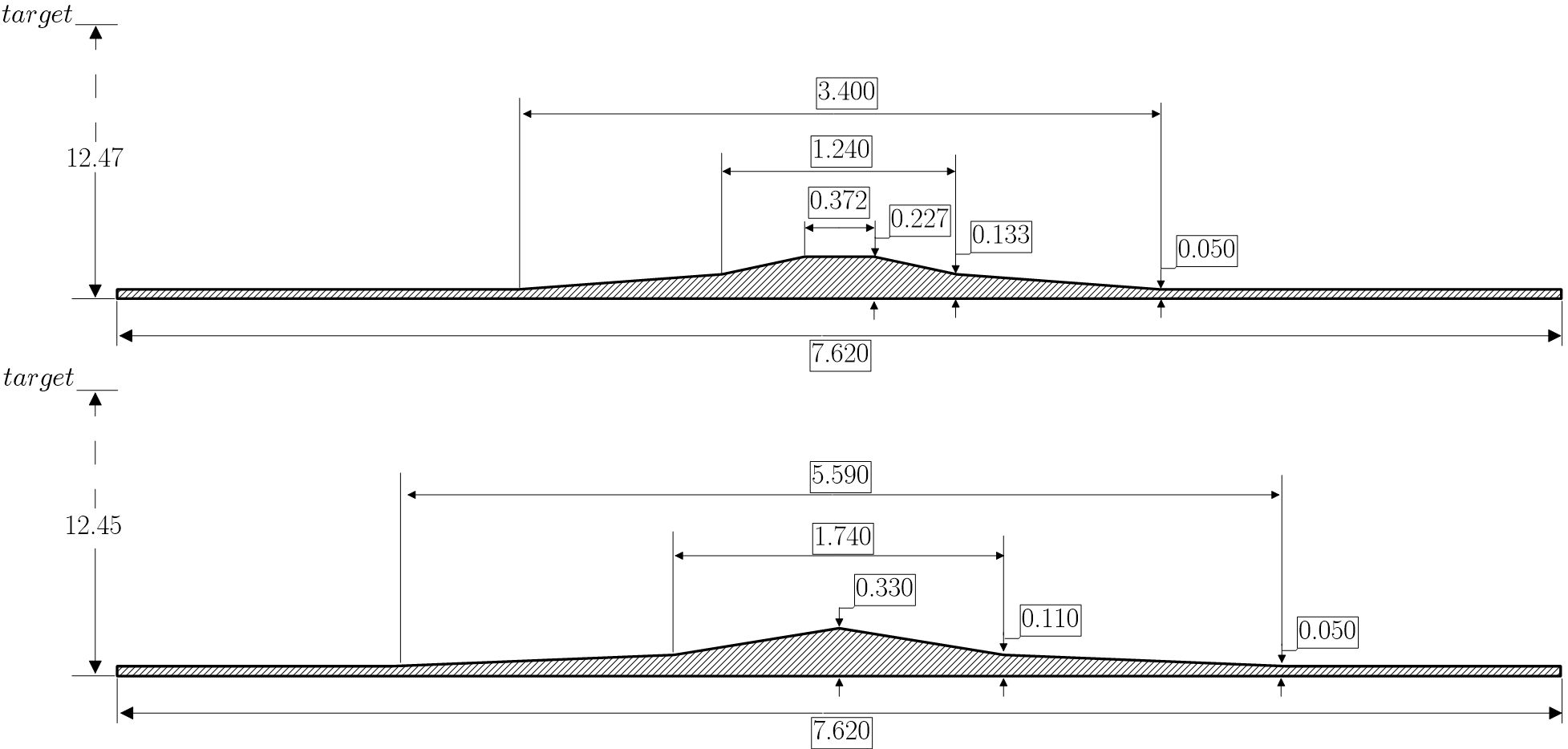}
\caption{Cross section of the {\it ad hoc} 6-FFF (above) and 10-FFF (below) filters. The materials are bronze and tantalum, respectively. Cylindrical symmetry applies. All dimensions are given in centimeters.}
\label{fig:filters}
\end{center}
\end{figure*}

According to the manufacturer (information provided to TrueBeam users under non-disclosure agreement), there is a large similarity between the head assembly geometries of the Clinac~2100 and the TrueBeam linacs downstream of a plane situated just above the jaws. Thus, in the FakeBeam geometry modifications on the head assembly of the Clinac 2100 linac were introduced in the part upstream of the jaws. A trial and error process was used to find the shape of thin filters which, situated approximately at the location of the standard flattening filters of the Clinac~2100 machine, reproduce the dose distribution produced by the FFF beams of the TrueBeam linac in a water phantom. We shall name these filters 6-FFF and 10-FFF filters. The detailed geometry of these filters is shown in figure~\ref{fig:filters}. The material composition of the 6-FFF and 10-FFF filters is bronze (70\% copper, 30\% zinc, $\rho=$8.412~g/cm$^3$) and tantalum ($\rho=$16.654~g/cm$^3$), respectively. The selection of these materials was inspired by the composition found in other flattening filters from Varian. A minor change was made to the original geometry of the Clinac~2100, so that the beryllium window was repositioned at the upstream plane of the primary collimator. Consequently, only the target is at vacuum in the FakeBeam geometry. The FakeBeam geometry includes the modifications to the primary collimator and the secondary collimator (lead shield) proposed by Chibani and Ma for the Clinac 2100 \cite{Chibani2007}.

\subsection{Monte Carlo simulations}\label{sec:MC}
All the simulations included in this work were performed with the PRIMO code~\cite{Rodriguez2013}, a user-friendly and freely--distributed software (http://www.primoproject.net) specifically designed for simulating Varian and Elekta linacs and estimating absorbed dose distributions in water phantoms and computerized tomographies. PRIMO is designed with a layered structure having the Monte Carlo general-purpose radiation transport code {\sc penelope}~\cite{PENELOPE2011} and the geometry packages {\sc penVox} and {\sc pengeom} as the lowermost layers. The main program used in PRIMO for steering the {\sc penelope} radiation transport subroutines is {\sc penEasy}~\cite{Sempau2011}. The modified geometry of the Clinac~2100 including the 6- and 10-FFF filters were coded as a hierarchic structure of homogeneous bodies limited by quadric surfaces according to the rules of the {\sc pengeom} package and were incorporated into PRIMO as a new linac named FakeBeam.

The \textsc{penelope} absorption energies ({\tt EABS}) were set to 100~keV for charged particles and 20~keV for photons everywhere. For bremsstrahlung emission by electrons and positrons~\cite{Llovet2005}, the energy of the emitted photons is sampled from the scaled cross-section tables of Seltzer and Berger~\cite{Seltzer1985} and its angular distribution is described by an approximation to the partial-wave shape functions of Kissel, Quarles and Pratt~\cite{kissel1983}. \textsc{penelope} uses a mixed simulation scheme \cite{Berger1963} for electrons and positrons and  detailed simulation for photons. The five \textsc{penelope} transport parameters for mixed simulation are {\tt C1}, {\tt C2}, {\tt WCC}, {\tt WCR} and {\tt DSMAX}. {\tt C1} and {\tt C2} are the electron average angular deflection and the maximum fractional energy loss allowed in one step, respectively. {\tt WCC} and {\tt WCR} are the energy cutoffs for inelastic and bremsstrahlung interactions, respectively. {\tt DSMAX} is the maximum allowed step length. In all simulations the condensed history transport parameters were set to {\tt C1}$=${\tt C2}$=$0.1, {\tt WCC}$=$100~keV and {\tt WCR}$=$20~keV for all materials, except for the materials composing the target where they were set to {\tt C1}$=${\tt C2}$=$0.001, {\tt WCC}$=$1~keV and {\tt WCR}$=$20~keV. We have observed that those parameters are adequate for producing an accurate transport in the target~\cite{Rodriguez2015}. The maximum step length ({\tt DSMAX}) was set to $1/10$ of the thickness of each material body. The initial electron source for the nominal 6~MV beam had a Gaussian energy distribution with mean of 5.8~MeV and full width at half maximum (FWHM) of 1\%. Its spatial distribution was also modeled by a Gaussian function centered at the linac central axis with a FWHM of 0.15~cm. For the nominal 10~MV beam the initial electron source was monoenergetic with an energy of 10.8~MeV and spatially distributed by a Gaussian function with a FWHM of 0.1~cm. These parameters were found optimal to match the measurements.

Several variance-reduction techniques were employed. The splitting-roulette~\cite{Rodriguez:2012ex}, which is a technique that combines photon splitting and Russian roulette applied to photons and electrons, was used in the target. The moveable skins technique~\cite{Brualla2009} was applied in the jaws and the primary collimator. Moveable skins is a particular implementation of range rejection in which external layers of high atomic number collimation structures are geometrically separated from the internal regions. These layers are called `skins'. In skins, all particles undergo analogue simulation ({\it i.e.}, no variance-reduction technique is applied), whereas in internal regions electrons are terminated by setting arbitrarily high energy cutoffs. Additionally, simple splitting was applied to all particles crossing the surface of the water phantom. The average standard statistical uncertainty of all bins scoring more than 50\% of the maximum dose was 0.3\% for all simulations of the 10 MV FFF beam and ranged between 0.4\% and 0.6\% for simulations of the 6 MV FFF beam. The $40\times40$~cm$^2$ field was used to iteratively tune the model and the rest were used to validate it.

Additionally, simulations employing the second generation of the Varian distributed TrueBeam phase-space files as sources of particles were also run. The phase-space files that belong to the second generation are those tallied on a plane, while those that belong to the first generation were tallied on a section of a cylindrical surface and employed a substantially lower number of primary particles~\cite{Constantin2011}. The phase-spaces were imported into PRIMO. The region downstream of the jaws was simulated as described above for the FakeBeam geometry and following the Varian recommendations. Splitting in the water phantom was used with a variable splitting factor depending on the field. The factors were selected to be sufficiently large to approximate the dose uncertainty to the latent variance of the phase space~\cite{Sempau2001}. The obtained average standard statistical uncertainty of all bins scoring more than 50\% of the maximum dose was about 0.8\%. We refer to these simulations as TrueBeam.

\begin{figure*}
\begin{center}
\includegraphics[scale=0.38]{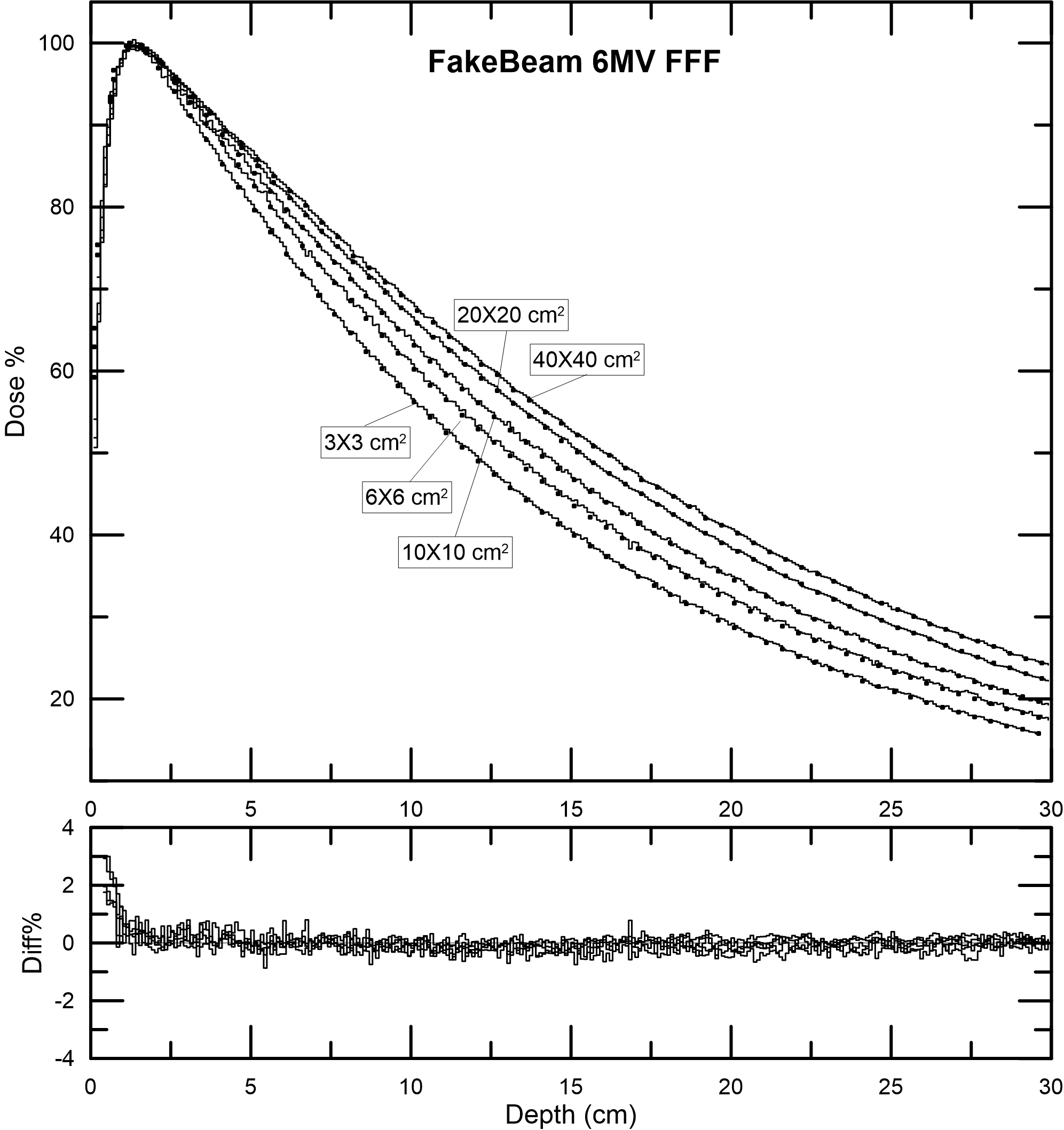}
\includegraphics[scale=0.38]{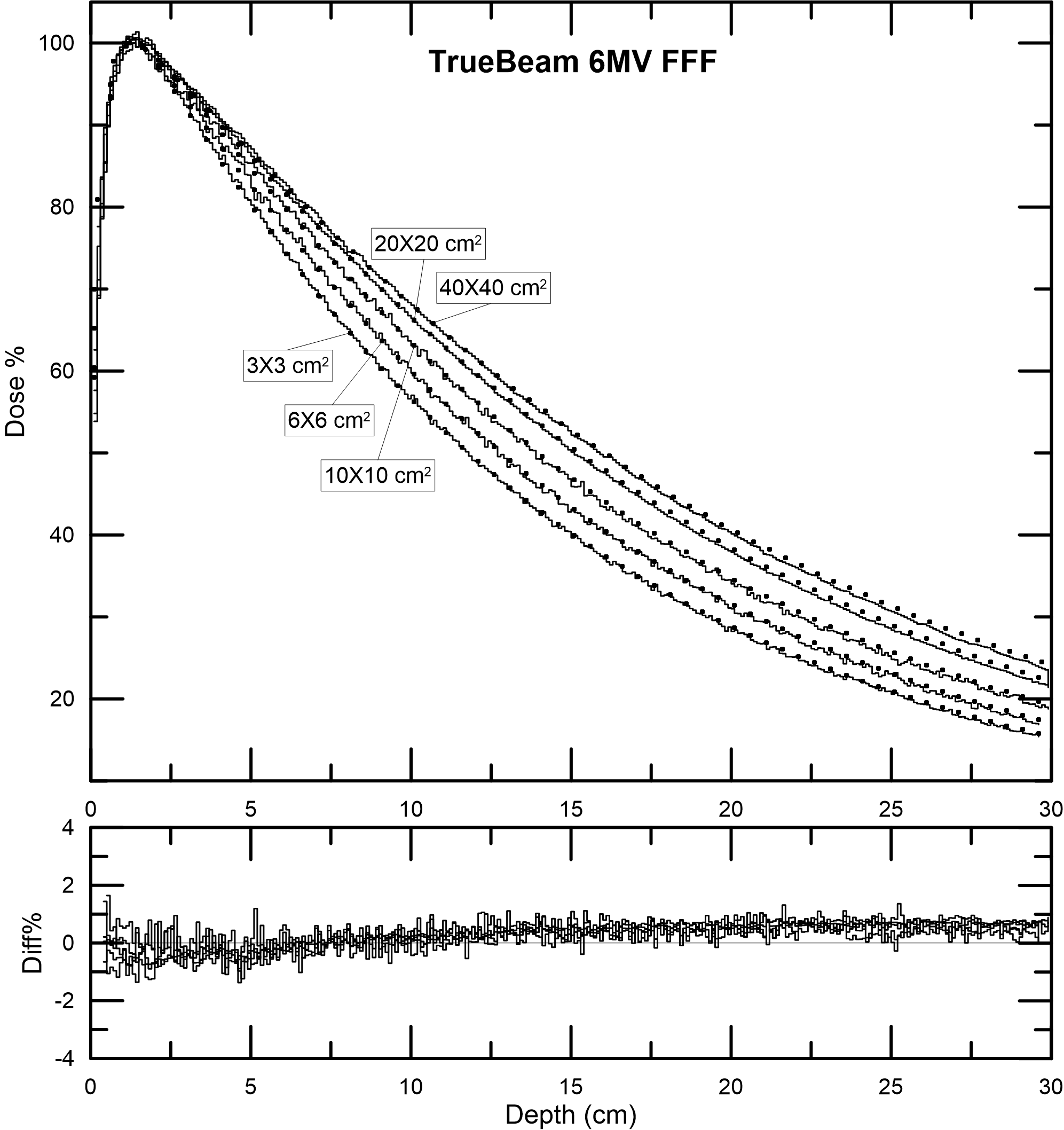}
\includegraphics[scale=0.38]{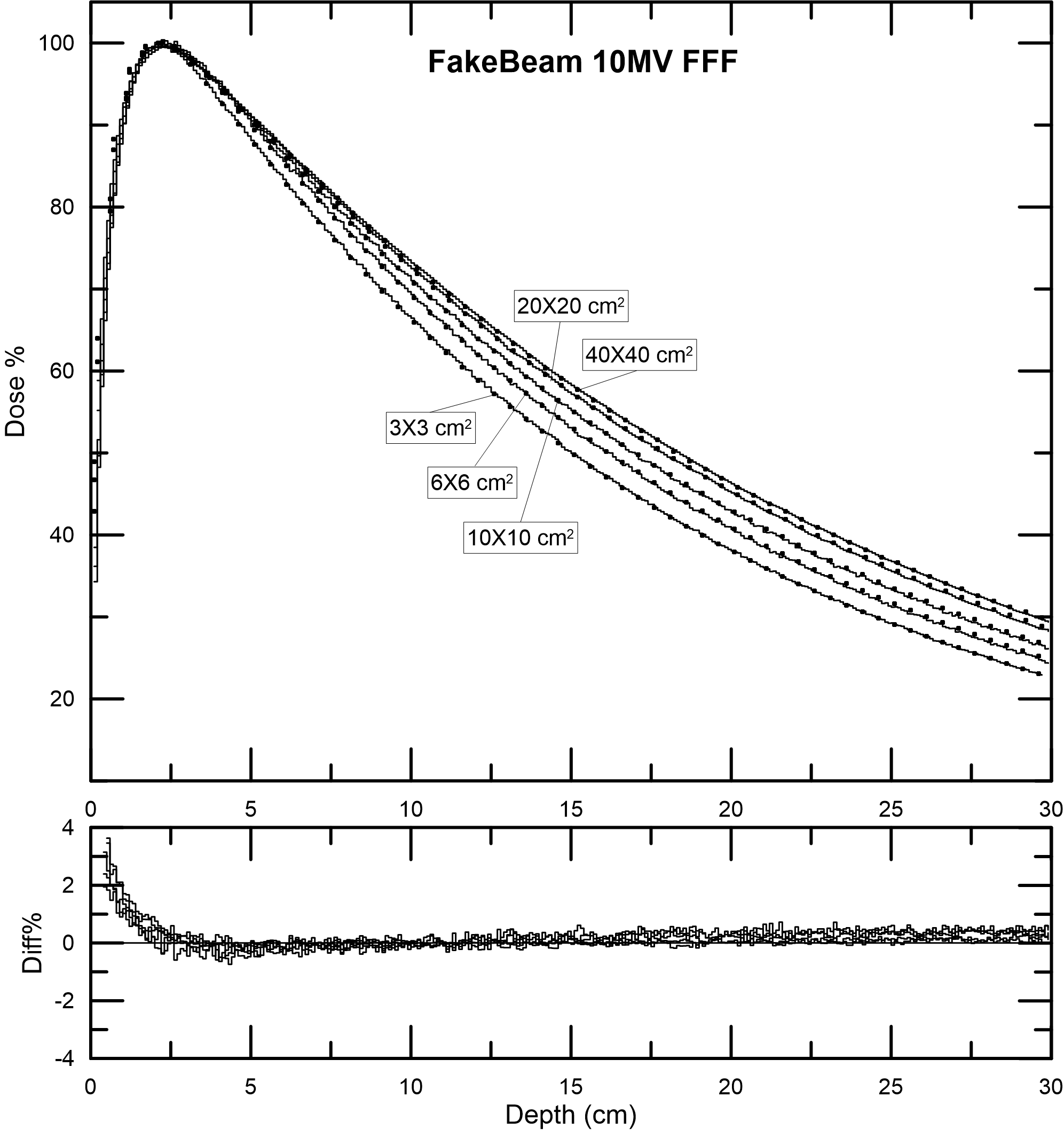}
\includegraphics[scale=0.38]{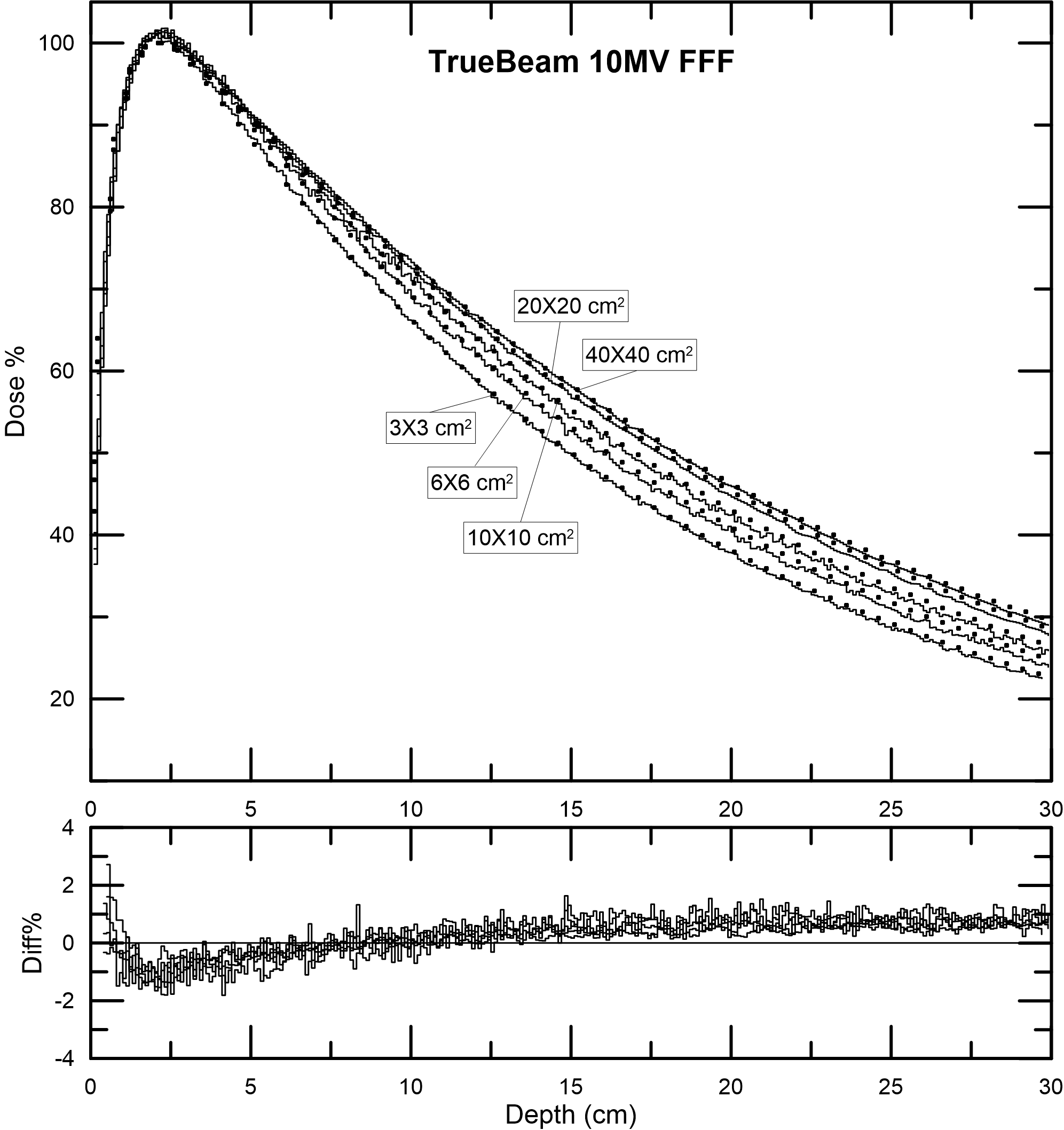}
\caption{Comparison of measured (dots) and Monte Carlo estimated (steps) depth-dose curves for the FakeBeam and TrueBeam geometries and the 6 and 10 FFF beams. Percentage dose differences are relative to the maximum dose. Statistical uncertainties are in the range of 0.3 to 0.8\% and are not shown.}
\label{fig:pdds}
\end{center}
\end{figure*}

\subsection{Geometry validation}\label{sec:val}
Validation of the geometry was based on comparisons between Monte Carlo estimated dose distributions and two sets of measurements: (i) the Varian Golden Beam Data Set (GBDS) which employed an IBA (Schwarzenbruck, Germany) CC13 ionization chamber; (ii) in-house measurements performed with an IBA PFD$^{3\tt{G}}$ detector \cite{IBAdiodes}. Comparisons were made for field sizes of 3$\times$3, 6$\times$6, 10$\times$10, 20$\times$20 and 40$\times$40~cm$^2$. Crossline profiles were compared at depth of maximum dose $d_{max}$, at 5.0~cm and at 10.0~cm depths. Measurements with the diode were used for comparing the depth-dose curves of the 3$\times$3~cm$^2$ field and lateral dose profiles of the 3$\times$3, 6$\times$6 and 10$\times$10~cm$^2$. The GBDS was employed for the rest of depth-dose and crossline dose profiles comparisons. The SSD was 100~cm and the phantom bin size was selected according to the field size: for the 3$\times$3, 6$\times$6 and 10$\times$10~cm$^2$ fields it was 0.1$\times$0.4$\times$0.1~cm$^3$; for larger fields it was 0.2$\times$0.8$\times$0.2~cm$^3$ (crossline$\times$inline$\times$depth). The larger bin size in the inline direction was chosen in order to reach a low uncertainty in the simulations with the TrueBeam phase-space files.

\begin{figure*}
\begin{center}
\includegraphics[scale=0.5]{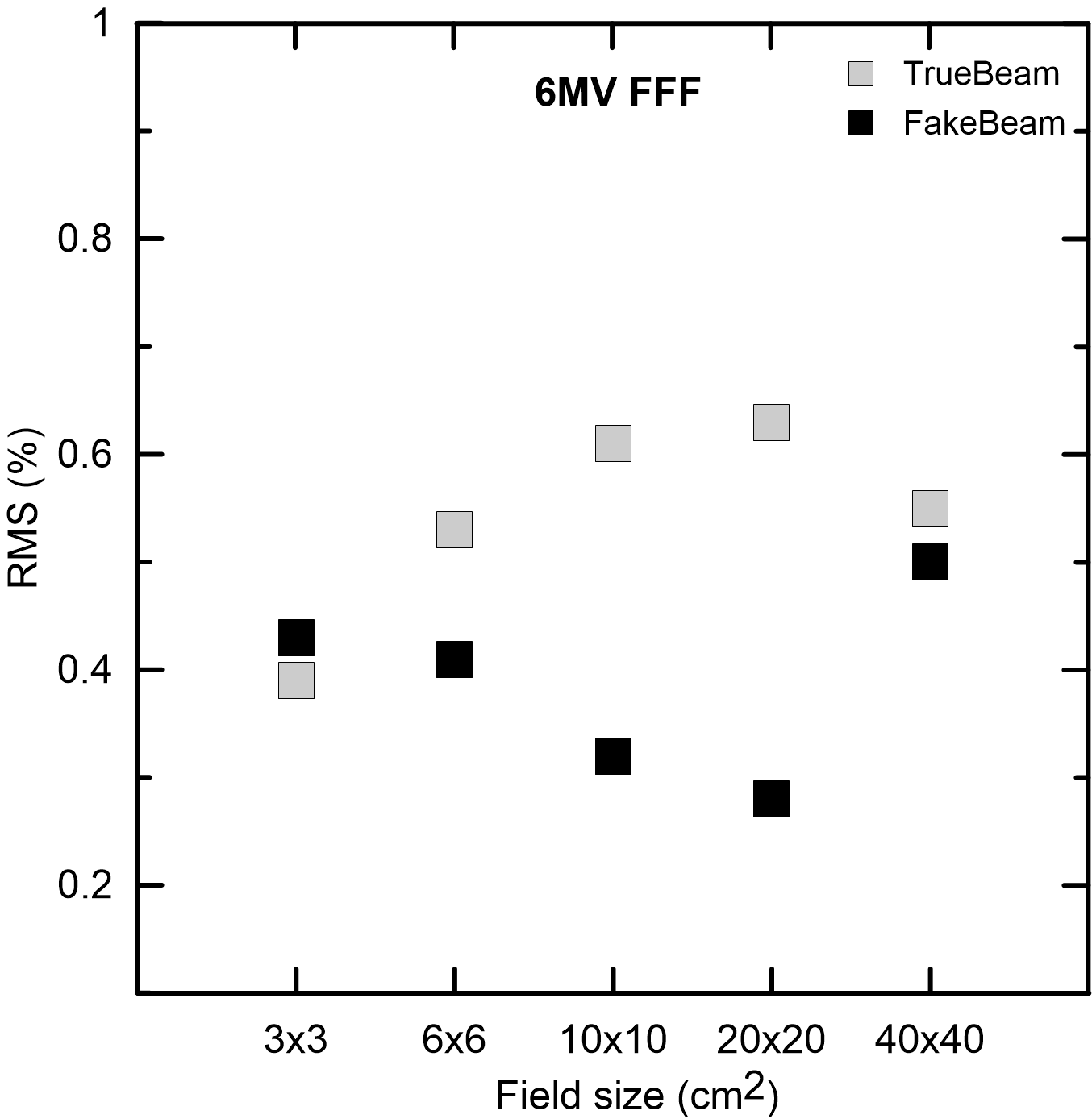}
\includegraphics[scale=0.5]{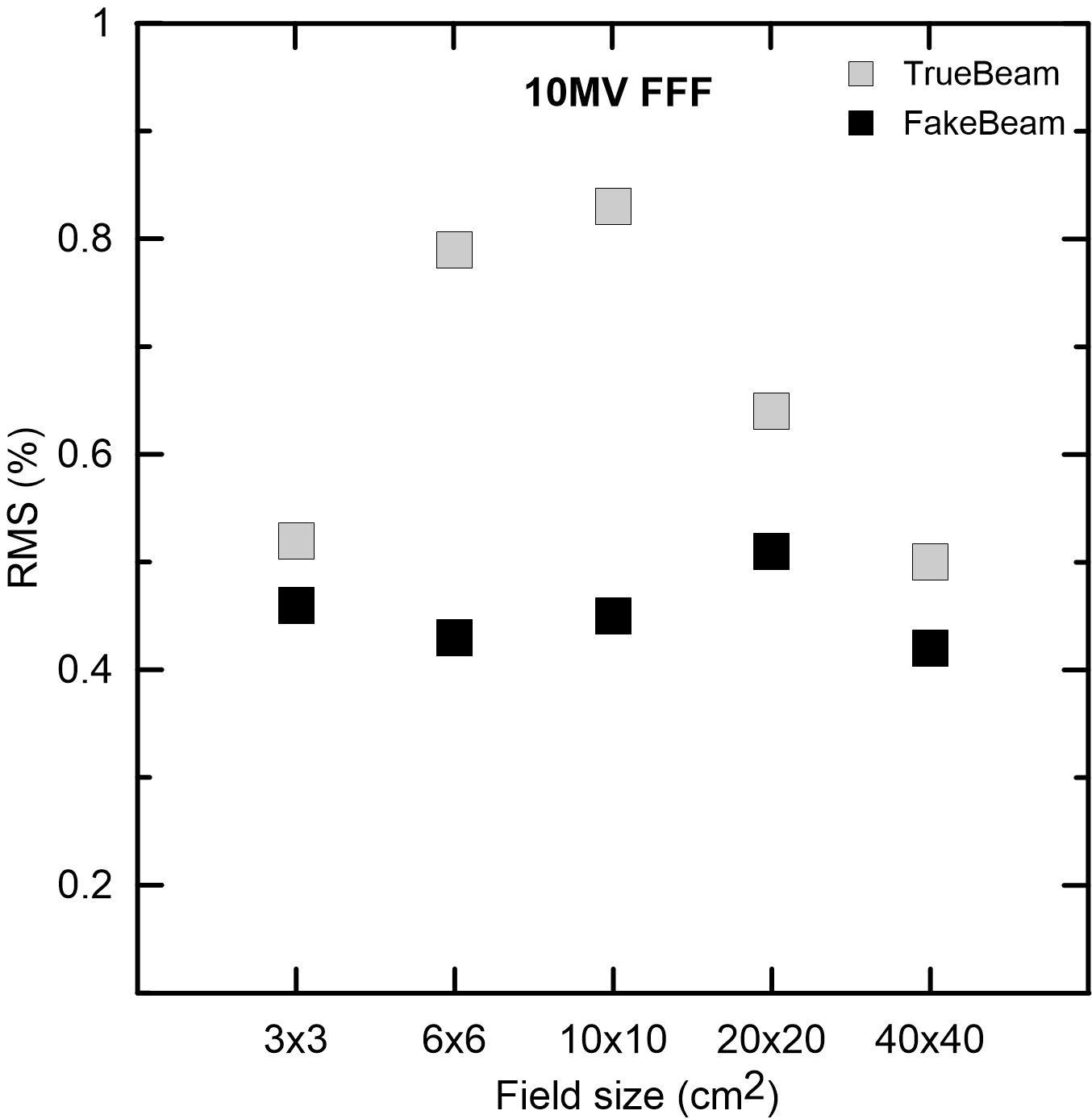}
\caption{Root-mean-square error of the dose difference between measured and Monte Carlo estimated depth-dose curves for the FakeBeam and TrueBeam geometries and the 6 and 10 MV FFF beams.}
\label{fig:rms}
\end{center}
\end{figure*}

\begin{figure*}
\begin{center}
\includegraphics[scale=0.35]{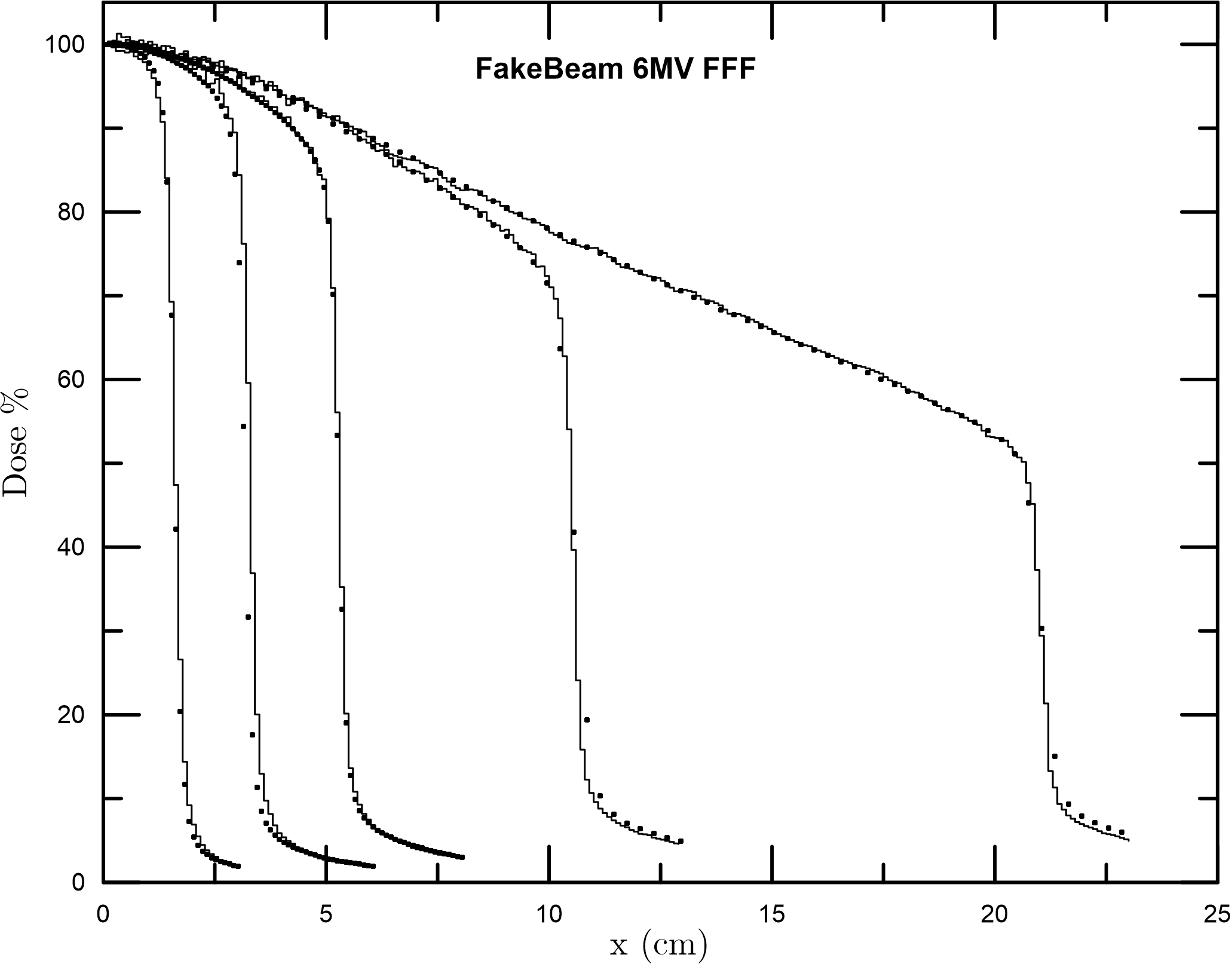}
\includegraphics[scale=0.35]{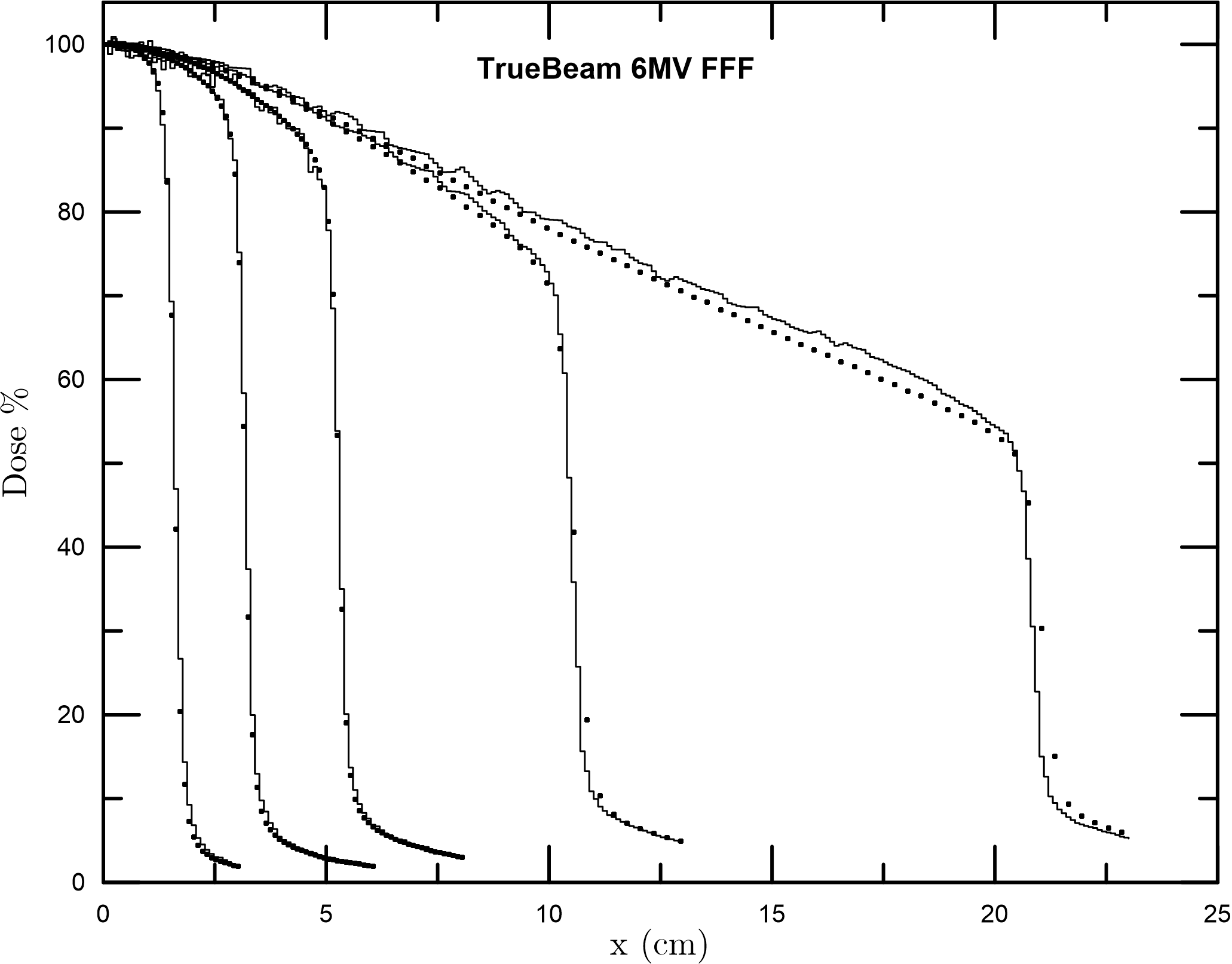}
\includegraphics[scale=0.35]{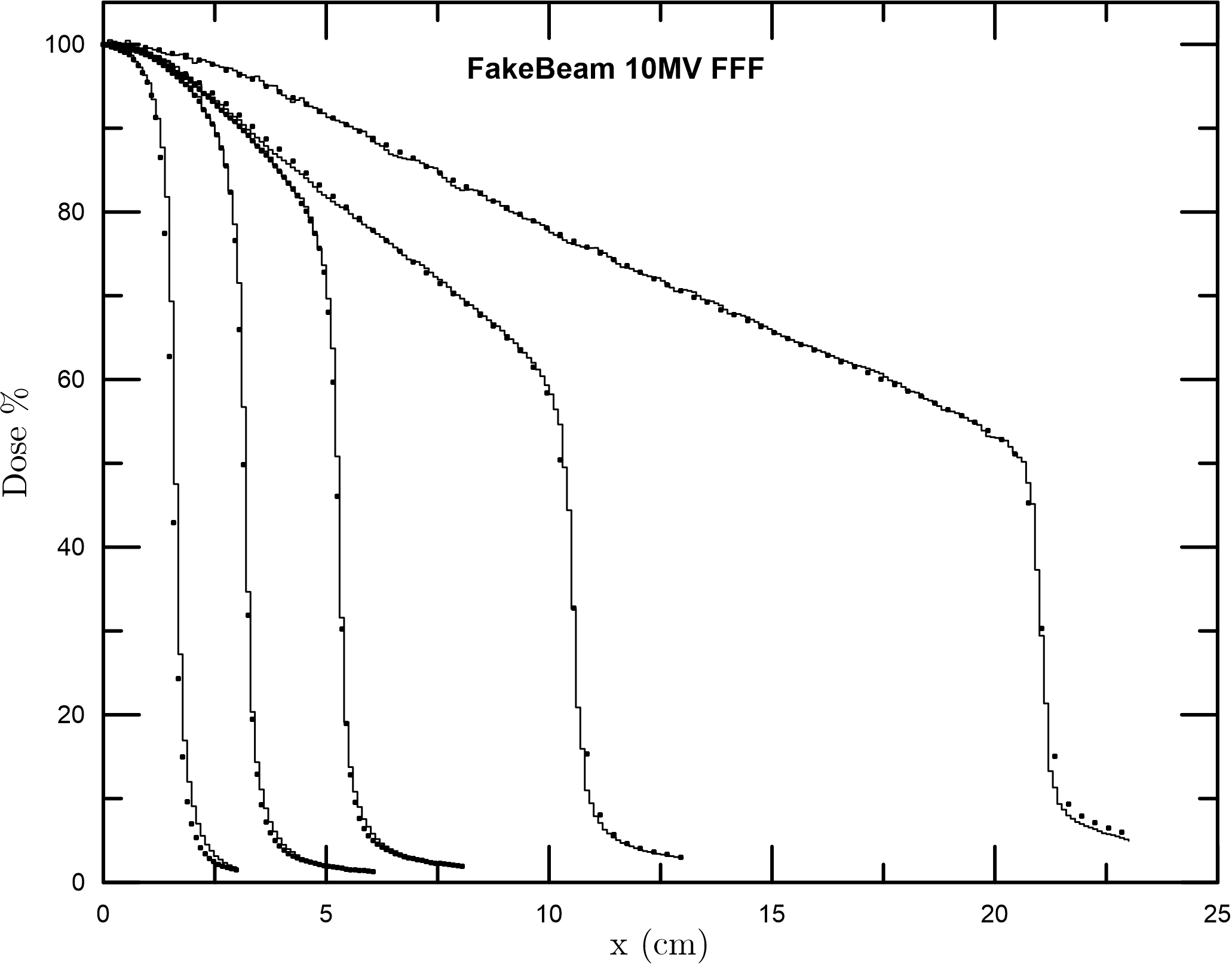}
\includegraphics[scale=0.35]{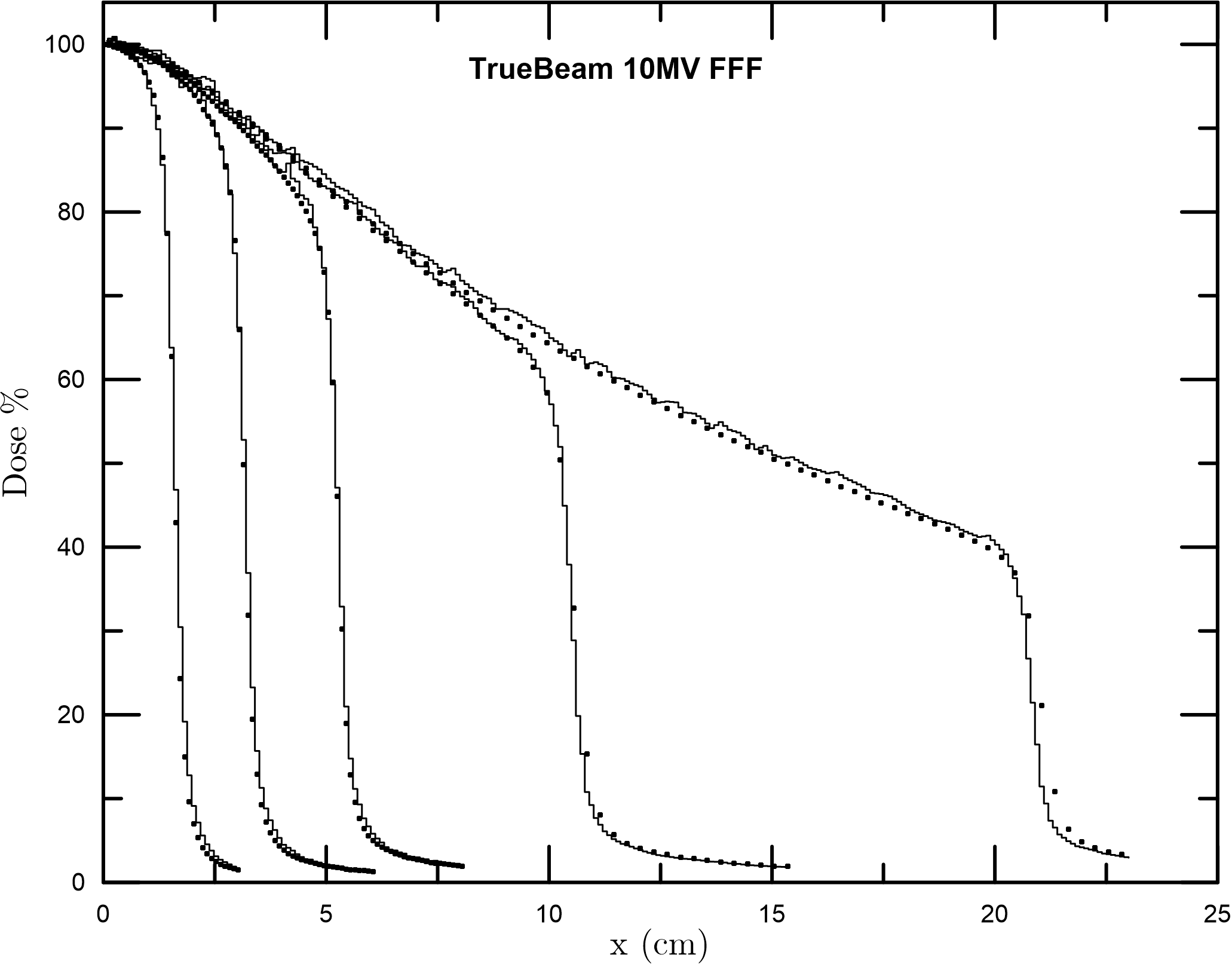}
\caption{Comparison of measured (dots) and Monte Carlo estimated (steps) crossline profiles at $5.0$ cm depth for the FakeBeam and TrueBeam geometries and the 6 and 10 MV FFF beams. Statistical uncertainties are in the range of 0.3 to 0.8\% and are not shown.}
\label{fig:profs}
\end{center}
\end{figure*}

The percentage of points passing a gamma analysis \cite{Low:1998wp} with criteria of 1\%, 1mm ($\gamma_{1,1}$), the dose difference relative to the maximum dose (diff\%) and the root-mean-squared error (RMS) of the dose difference were employed as figures of merit to determine agreement between the measured and the Monte Carlo estimated dose distributions. The RMS was computed only for depth-dose curves at depths greater than 3 mm. In spite of the fact that the reference data set in the gamma analysis is a 1D array of dose points with arbitrary separation, PRIMO performs a re-gridding process of the evaluated 3D dose distribution that allows to apply an arbitrarily small distance-to-agreement criterion. Re-gridding is performed using trilinear interpolation. In the same manner, the dose profiles used for evaluating the percentage difference were calculated in the Monte Carlo estimated 3D dose distribution by trilinear interpolation at the measurement points. In order to reduce the effect of uncertainties in the analysis, estimated depth-dose curves were normalized so that their areas between $d_{max}$ and $d_{50\%}$ coincided with the experimental values~\cite{Gete2013}. Crossline profiles were normalized to 100\% at the central axis.

\subsection{Sensitivity analysis}\label{sec:SA}
We performed a series of simulations with the purpose of assessing the sensitivity of the dose profiles produced by the FakeBeam model to the initial beam parameters of the beam. The depth-dose curves and lateral profiles at 5.0~cm of depth obtained with the initial parameters described in section \ref{sec:MC} for a 40$\times$40~cm$^2$ field were used as reference in the gamma analysis. For the 6~MV beam we varied the initial energy of electrons from 5.6~MeV to 6.0~MeV in steps of 0.1~MeV. The FWHM of the focal spot size was varied from 0 to 0.15~cm in steps of 0.05~cm. For the 10~MV beam the initial energy ranged from 10.6 to 11.0~MeV with a step of 0.2~MeV and the FWHM of the focal spot size was varied from 0 to 0.1~cm in steps of 0.05~cm. The FWHM of the initial energy was set to 0 in all cases. Additionally, we performed simulations with the initial beam parameters that were used to generate the Varian's phases-space files. These parameters are declared in the phase-space file headers and are, for the 6~MV beam, 5.9~MeV, 0.06~MeV and 0.078~cm for the initial energy, FWHM of the energy and FWHM of the focal spot size, respectively. Likewise, for the 10~MV beam they were 10.2~MeV, 0.102~MeV and 0.096~cm. The obtained average standard statistical uncertainty of the dose distributions varied between 0.5\% and 1\%.

\begin{figure*}
\begin{center}
\includegraphics[scale=0.35]{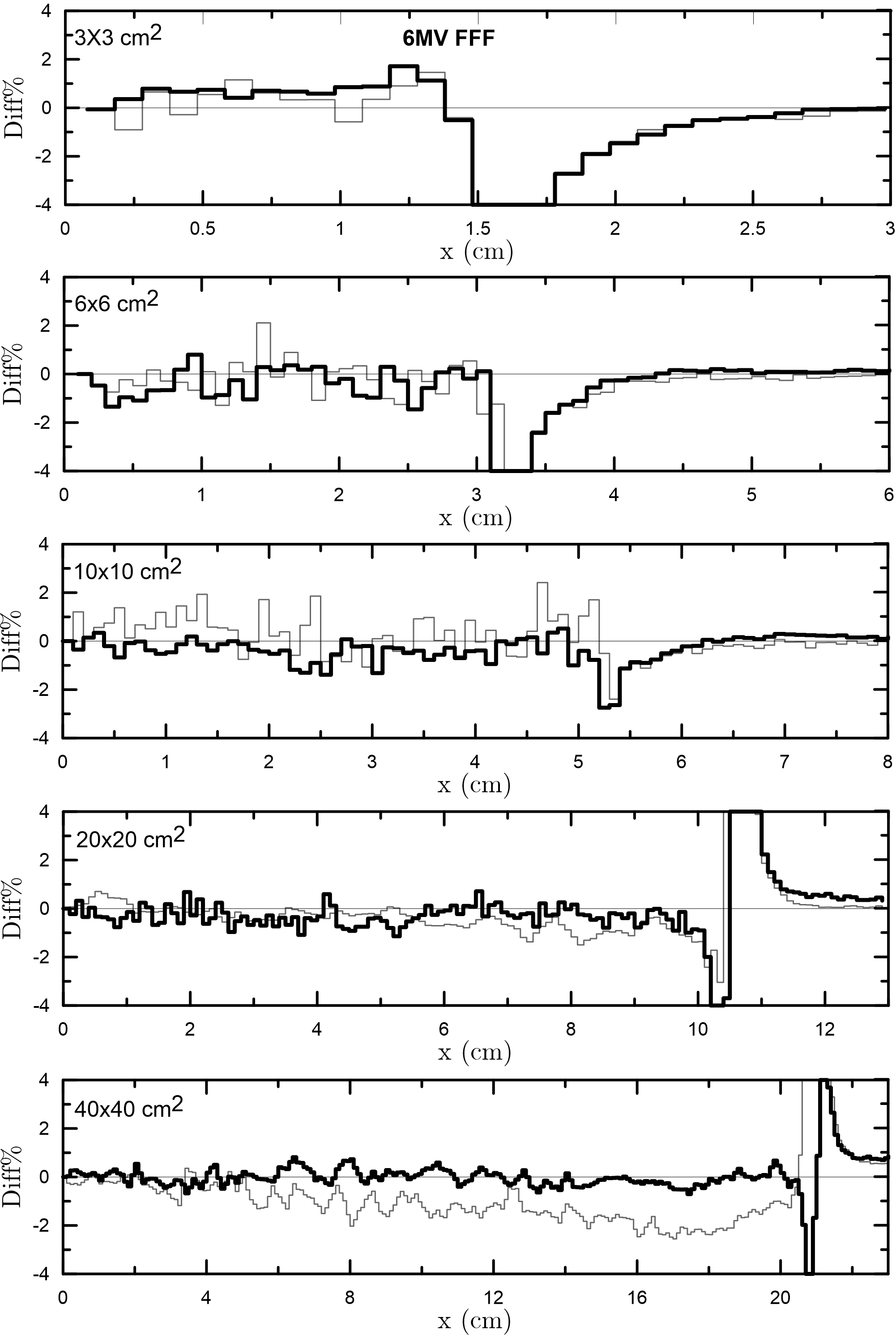}
\includegraphics[scale=0.35]{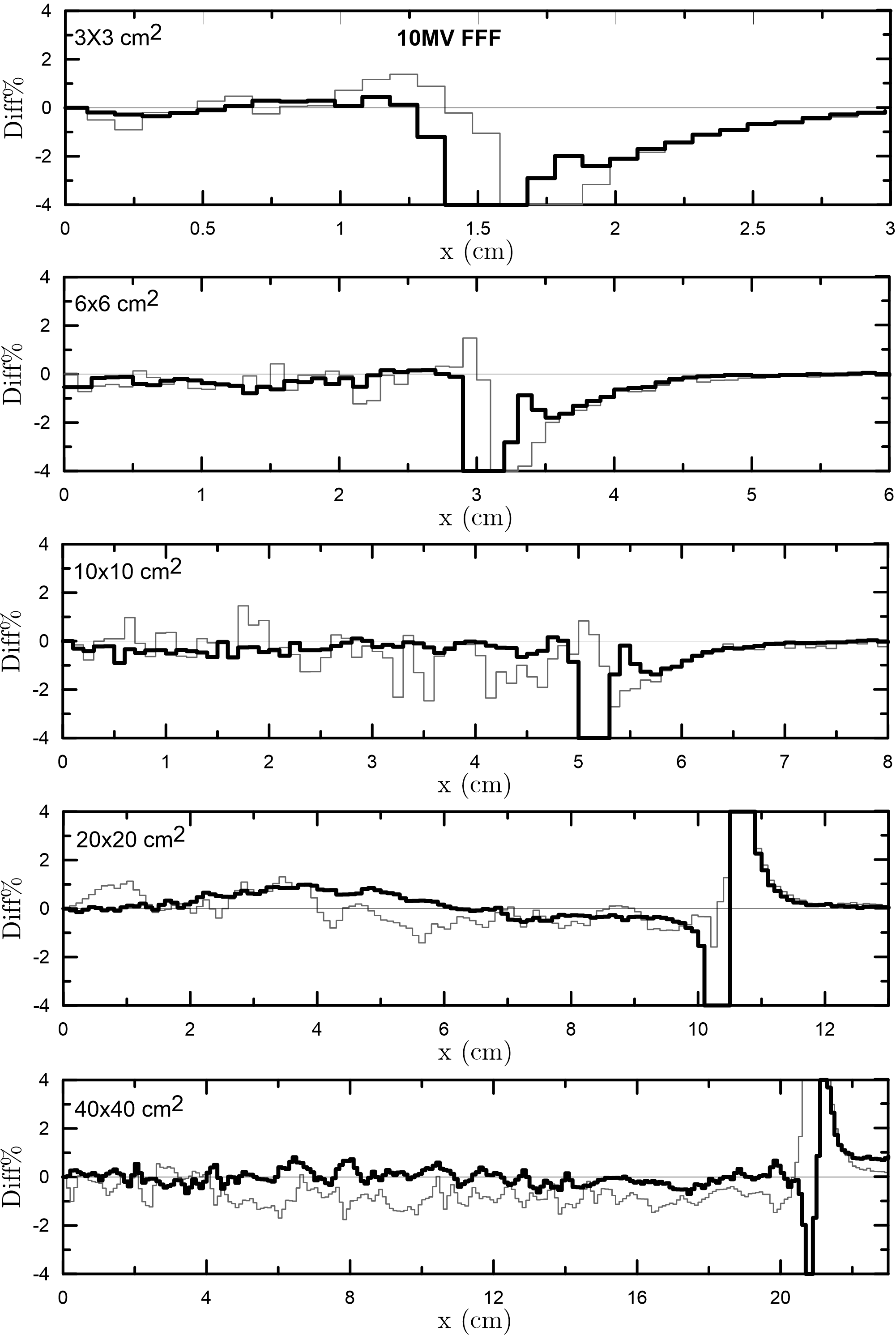}
\caption{Relative difference between measured and Monte Carlo estimated lateral profiles at 5 cm of depth for the FakeBeam (black) and TrueBeam (gray) geometries for the 6 (left) and 10 (right) MV FFF beams.}
\label{fig:pdiff}
\end{center}
\end{figure*}

\section{Results}\label{sec:Res}
Figure \ref{fig:pdds} shows the comparison between measured and Monte Carlo estimated depth-dose curves. For FakeBeam, differences are less than 1\% in most of the points for both beams and all fields evaluated, except in the high dose gradient build-up region, near the phantom surface, where an over-response of the cylindrical ionization chamber is expected. This over-response arises from the fact that ionization chambers, even those with a small cavity volume such as the CC13 employed for measuring the Varian's GBDS, overestimate the dose in the build-up region where lack of charged particle equilibrium exists. Depth-dose curves for TrueBeam also show a good agreement for both beams with lower differences in the build-up region than FakeBeam but slightly larger beyond $d_{max}$, which results in relatively lower RMS values for FakeBeam as shown in figure~\ref{fig:rms}. Values of $\gamma_{1,1}$ obtained were in all cases, for both the FakeBeam and the TrueBeam simulations, greater than 98\%.

\begin{figure}
\begin{center}
\includegraphics[scale=0.5]{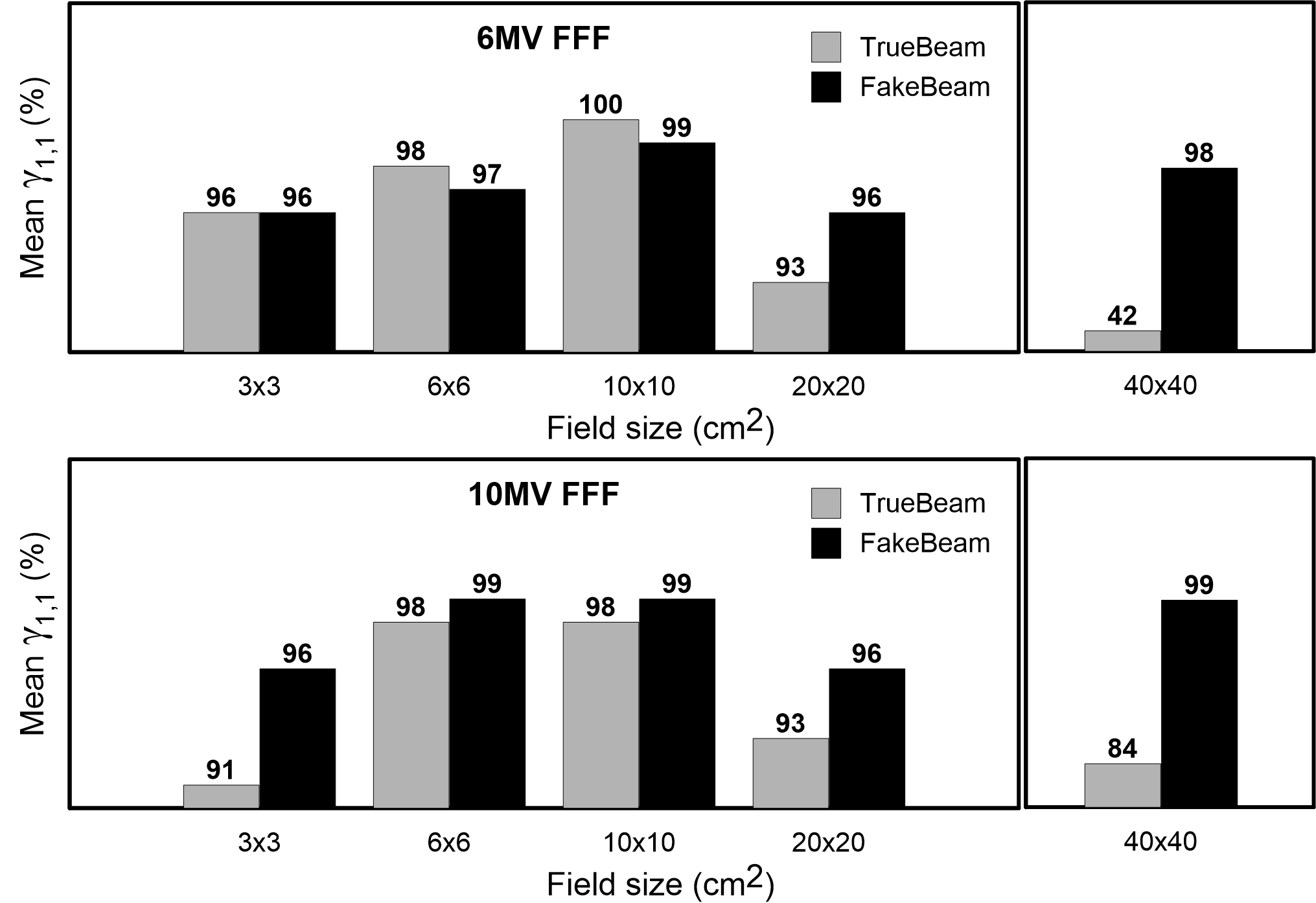}
\caption{Results of gamma analysis of lateral dose profiles. Bars represent the values of the combined $\gamma_{1,1}$ for the three depths considered in the analysis. Notice that results for the $40\times40$~cm$^2$ field are shown in a different scale.}
\label{fig:gamma}
\end{center}
\end{figure}

Comparison of lateral dose profiles shown in figures \ref{fig:profs} and \ref{fig:pdiff} reveals a good agreement of FakeBeam with the experimental data with differences less than 1\% for large fields ($20\times20$~cm$^2$ and $40\times40$~cm$^2$) and less than 1.5\% for smaller fields, except in the regions of high dose gradient. These differences are similar to those found for TrueBeam for the smaller fields but lower for large fields where TrueBeam agreement is poorer, particularly for the 6 MV FFF beam.

Values of the combined $\gamma_{1,1}$ obtained for the lateral dose profiles at the three depths considered are shown in figure \ref{fig:gamma}. The agreement is very good for FakeBeam with at least 96\% of the points passing the test for all fields and energies. Gamma analysis also reveals the poor agreement obtained with the phase-space files distributed by Varian for TrueBeam for the $40\times40$~cm$^2$ field with 42\% and 84\% of the points passing the test, for the 6 MV and 10 MV beams, respectively. These values improved to 88\% and 98\% when the gamma analysis criteria were relaxed to 2\%, 2~mm.

Results of the sensitivity analysis show that the dose is quite insensitive to the initial parameters of the beam in the range analyzed for both nominal energies. Dose profiles and depth dose curves matched the reference data sets with $\gamma_{1,1}(\%)$ results of 98\% or better. For the lateral dose profiles obtained with the parameters used to generate the Varian's phase-space files, the 6~MV beam produced a $\gamma_{1,1}$ and $\gamma_{2,2}$ of 97\% and 100\%, respectively. In contradistinction, for the 10~MV beam the $\gamma_{1,1}$ and $\gamma_{2,2}$ were 53\% and 95\%, respectively. This larger variation is caused by the marked difference between the initial energy of the reference and the analyzed beams, which is of about 0.5-0.7~MeV.

\section{Summary and Conclusions}\label{sec:Conclusions}

A geometrical model, named FakeBeam, for the Monte Carlo simulation of FFF beams produced by the TrueBeam linac has been developed and validated by comparison of computed PDDs and profiles with experimental measurements.  The description of the filters developed in this work allows those researchers having access to the Varian Clinac 2100 geometry to implement FakeBeam in their own Monte Carlo systems. FakeBeam has been incorporated into PRIMO as a new linac model, so users can simulate it without any coding effort. PRIMO includes the geometries of the Millennium 120 and 120 HD multileaf collimators. Users of the PRIMO code can either simulate the FakeBeam geometry or import the Varian's distributed phase-space files for TrueBeam. Our approach overcomes some of the limitations of using phase-space files as a source of particles. Now it is possible to: (i) adapt the initial beam parameters to match measured dose profiles; (ii) reduce the statistical uncertainty to arbitrary low values; and (iii) assess systematic uncertainties (type B) by employing different Monte Carlo codes. Our model does not take into account the actual geometry of the ionization chamber of the TrueBeam linac. Instead, it uses the ionization chamber of the Varian Clinac 2100. Since knowledge of the ionization chamber geometry is essential for some methods of Monte Carlo absolute dosimetry, our model is limited in this regard. Notwithstanding, the FakeBeam geometry could be used in conjunction with the method employed by Zavgorodni {\it et. al} \cite{Zavgorodni2014} to determine backscattered radiation in the monitor chamber. With this combined approach it is possible to provide dose distributions expressed in Gy/MU with an accuracy close to 1\%.

\section*{Acknowledgments}\label{sec:ack}
MR, JS and LB are thankful to the Spanish Ministerio de Econom\'{\i}a y Competitividad (project FIS2012-38480) and to the Deutsche Forschungsgemeinschaft (project BR~4043/3-1). JS is grateful to the Spanish Networking Research Center CIBER-BBN for financial support.

\end{document}